\documentclass{article}

\usepackage{amsmath,amssymb,a4wide}
\usepackage{epsfig}

\numberwithin{equation}{section}

\title{Scaling function for self-avoiding polygons}

\author{C.~Richard$^{1)}$, I.~Jensen$^{2)}$ and A.~J.~Guttmann$^{2)}$\\
\\
${}^{1)}$Institut f\"ur Mathematik, Universit\"at Greifswald\\
Jahnstr. 15a, 17487 Greifswald, Germany\\
\\
${}^{2)}$Department of Mathematics and Statistics\\
University of Melbourne, Victoria 3010, Australia}

\begin{document}

\maketitle

\begin{abstract}
Exactly solvable models of planar polygons, weighted by perimeter and
area, have deepened our understanding of the critical behaviour of
polygon models in recent years. 
Based on these results, we derive a conjecture for the exact form of
the critical scaling function for planar self-avoiding polygons.
The validity of this conjecture was recently tested numerically using
exact enumeration data for small values of the perimeter on the square
and triangular lattices. 
We have substantially extended these enumerations and also enumerated
polygons on the hexagonal lattice. 
We also performed Monte-Carlo simulations of the model on the square lattice. 
Our analysis supports the conjecture that the scaling function is
given by the logarithm of an Airy function.
\end{abstract}

\section{Introduction}

The model of self-avoiding polygons (SAPs) is an important unsolved
combinatorial problem in statistical mechanics \cite{MS93,Ja00}.
In this article, we consider the model of SAPs in two dimensions. 
SAPs counted by perimeter are the canonical model of ring polymers,
while when also counted by area, they serve as a model of vesicles
\cite{LSF87}.
The model also has connections with many other models of statistical
mechanics where boundaries of certain cluster types play a role,
notably in percolation and in spin models such as the Ising model and
the Potts model, see also \cite{CZ02}.
Despite its importance, there are only a few known rigorous results
about SAPs, almost all information arising from numerical
investigations \cite{FGW91,MS93,FGW91,Ja00,CG93,C94}. Promising new
developments however arise from the theory of stochastic processes
\cite{LSW02}, see also \cite{C02} (this volume).
Recently, we verified numerically a conjecture of one of us about the
scaling function for SAPs, which describes the singular part of the
SAP perimeter and area generating function about its tricritical point
\cite{RGJ01,R02}.
The prediction of the scaling function also gives the distribution of
area moments.
The conjecture was already stated in 1995 \cite{PO95b}, as a result of
a systematic investigation of simple exactly solvable subclasses of
SAPs.
There it was noticed that the critical exponents of {\it rooted} SAPs
coincide with the exponents of staircase polygons, where the scaling
function had been computed \cite{P95}.
This led to the question whether both scaling functions are of the same type.

Over the years, it turned out that the mathematical structure
underlying exact solvability is given by $q$-algebraic functional
equations.
They arise from the requirement that the polygon model is built up
recursively, see e.g. \cite{Bou96}.
This general type of functional equation appears first, within the
framework of algebraic languages, in \cite{D99}, together with the
discussion of the area distribution for simple polygon models showing
a square-root singularity.
However, the article \cite{D99} does not contain a derivation of its
central result.
Recently, we investigated the scaling behaviour of general
$q$-algebraic functional equations \cite{RGJ01, R02}.
The techniques used for handling $q$-algebraic functional
equations are applicable in a wide variety of problems, see
e.g. \cite{K02} for exact low order cluster moments of asymmetric
compact directed percolation.
An important outcome of our previous analysis was that the scaling
behaviour generically depends only on the singularity of the perimeter
generating function.
This remarkable kind of universality further supported the above
conjecture and led us to test it numerically.

Very recently, investigations on the self-avoiding walk problem using
the theory of stochastic processes led to a number of predictions of
the behaviour of SAPs in the scaling limit.
One of them is that the scaling limit of the half plane infinite SAP
is the outer boundary of the union of two Brownian excursions from $0$
to $\infty$ in the upper half plane \cite{LSW02}.
Since the area under a Brownian excursion is Airy distributed
\cite{L84}, this would suggest that the area moments of SAPs obey an
Airy type distribution.
This is consistent with our conjecture that the SAP scaling function
is the logarithm of an Airy  function and also illuminates the connection with
the model of staircase polygons, whose scaling limit is described by 
Brownian excursions. 

In the remainder of this article we first introduce polygon models, in
particular self-avoiding polygons, and discuss their critical behaviour.
We then consider $q$-algebraic functional equations and show how to
extract scaling behaviour in the vicinity of a square-root singularity
from the functional equation.
The following section analyses the exactly solvable example of
bar-graph polygons, using the techniques described in the previous
section.
The next section on self-avoiding polygons contains a summary of the
analysis of recently obtained exact enumeration data for polygons on
the square lattice to perimeter 100, the hexagonal lattice to
perimeter 140 and the triangular lattice to perimeter 58.
The analysis confirms the conjectured form of the scaling function in
that predicted exact amplitude combinations are confirmed to within a
numerical accuracy of 5--6 significant digits.
This section also explains the Monte-Carlo simulation method, which we
performed on the square lattice for polygons up to perimeter 2048, and
confirms the conjectured form of the scaling function within a
numerical accuracy of 3 significant digits.
A concluding section discusses possible applications of our results
and methods.

\section{Polygon models and their scaling behaviour}

\begin{figure}[htb]
\begin{center}
\begin{minipage}[b]{0.3\textwidth}
\center{\epsfig{file=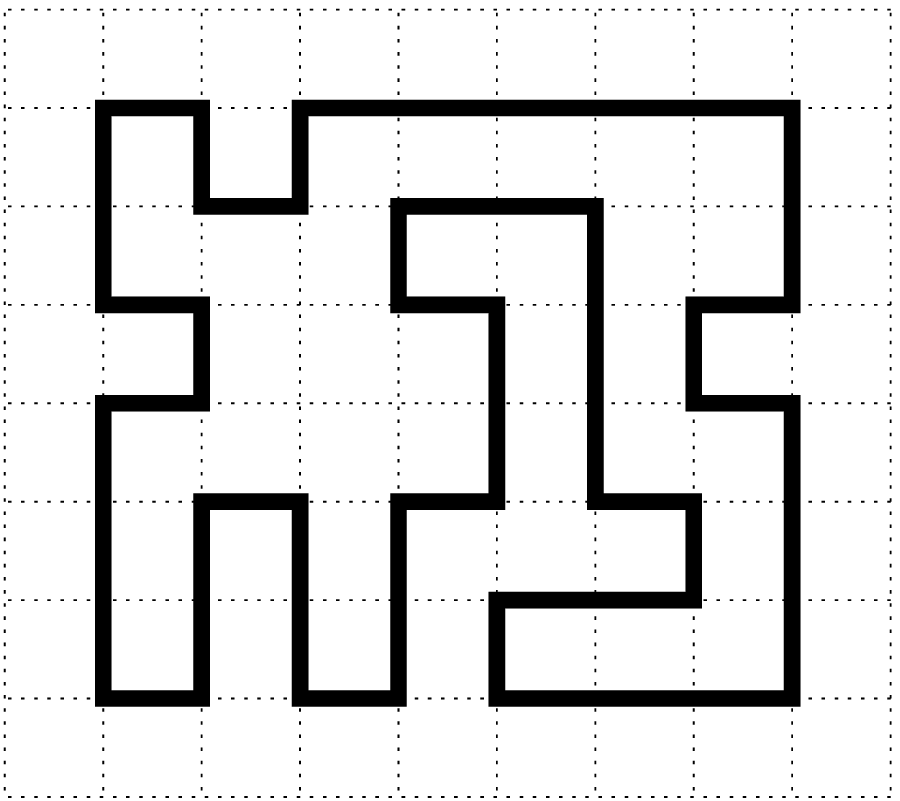,width=4cm}}
\end{minipage}
\hfill
\begin{minipage}[b]{0.3\textwidth}
\center{\epsfig{file=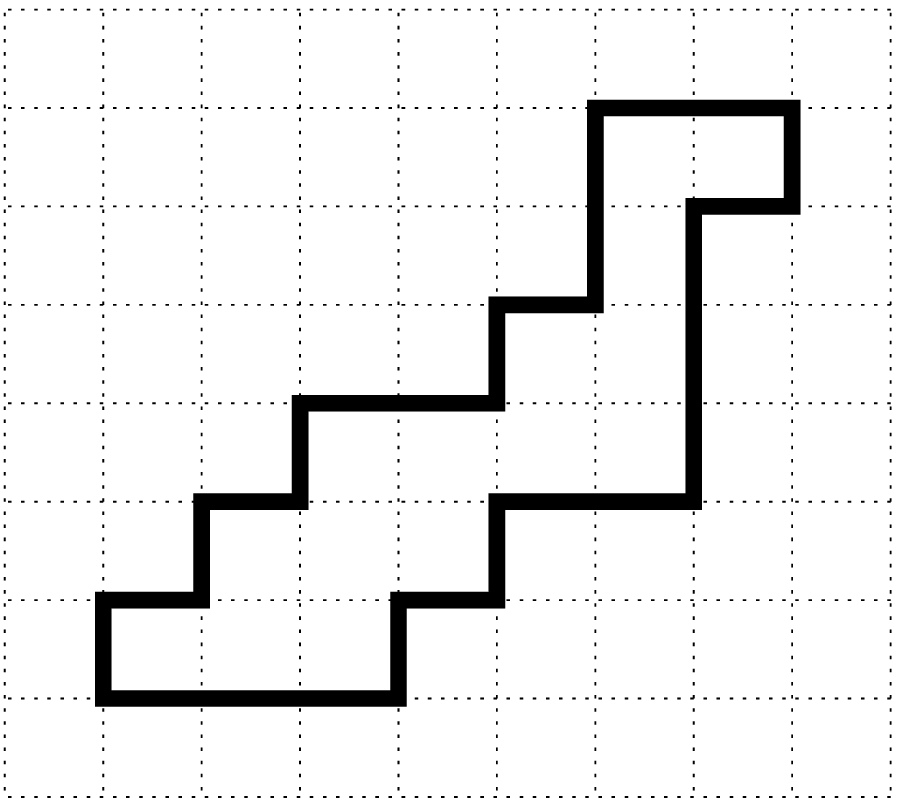,width=4cm}}
\end{minipage}
\hfill
\begin{minipage}[b]{0.3\textwidth}
\center{\epsfig{file=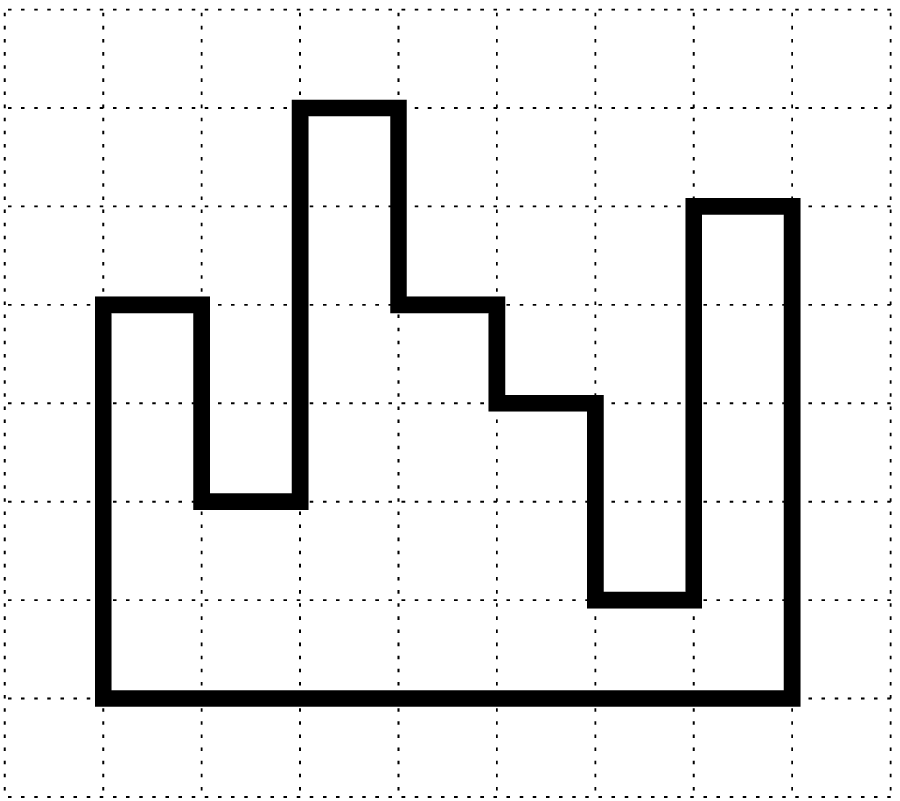,width=4cm}}
\end{minipage}
\end{center}
\caption{\label{fig:models}
\small Prominent polygon models. 
From left to right: self-avoiding polygons, staircase polygons,
bar-graph polygons.}
\end{figure}

For concreteness, we discuss models of polygons defined on the square
lattice $\mathbb Z^2$.
These models can be defined similarly for other lattices as well as in
the continuous case.
A {\it self-avoiding polygon (SAP)} is a closed, non-intersecting loop
on the edges of the square lattice.
Important subclasses are {\it staircase polygons} and {\it bar-graph
polygons} \cite{PB95}, whose obvious definition may be extracted from
Figure \ref{fig:models}.
Whereas the latter two models have been solved exactly, as also
discussed below, for SAPs there are very few rigorous results,
complemented by a number of numerical investigations.
Let $p_{m,n}$ denote the number of different polygons of perimeter $m$
and area $n$, respectively, where we regard translated polygons as
equal.
The object of interest is the {\it perimeter and area generating function}
\begin{equation}
G(x,q) = \sum_{m,n} p_{m,n} x^m q^n,
\end{equation}
where we introduced a perimeter activity $x$ and an area activity $q$.
Sometimes, horizontal and vertical perimeter are distinguished,
leading to an anisotropic perimeter and area generating function $G(x,y,q)$.
The function $G(x,1)$ is called {\it perimeter generating function}.
Of particular interest is the singular behaviour of $G(x,q)$ as a
function of $x$ and $q$, which translates to the behaviour of the
coefficients $p_{m,n}$ in the limit of large perimeter and area.

The above models display a so-called {\it collapse transition} between
a phase of extended and a phase of deflated polygons \cite{FGW91}:
in the extended phase $q=1$, the mean area of polygons $\langle a
\rangle_m$ of perimeter $m$ grows asymptotically like $m^{3/2}$,
whereas it grows like $m$ in the deflated phase $q<1$.
It can be shown that in the limit $q\to0$ the generating function is
dominated by polygons of minimal area.
Since for SAPs these polygons may be viewed as branched polymers, the
phase $q<1$ is also referred to as the {\it branched polymer phase}.
This change of asymptotic behaviour is reflected in the singular
behaviour of the perimeter and area generating function. 
Typically, the line $q=1$ is a line of finite essential singularities
for $x<x_c$.
The line $x_c(q)$, where $G(x,q)$ is singular for $q<1$, is typically
a line of algebraic (staircase, bar-graph) or logarithmic (SAPs)
singularity.
For branched polymers in the continuum limit, the logarithmic
singularity has been recently proved in \cite{BI01} (see also this
volume).

Of special interest is the point $(x_c,1)$ where these two lines of
singularities meet.
The behaviour of the singular part of the perimeter and area
generating function about $(x_c,1)$ is expected to take the special
form
\begin{equation}\label{form:scaling}
G(x,q) \sim G^{(reg)}(x,q) + (1-q)^\theta F((x_c-x)(1-q)^{-\phi}),
\qquad (x,q) \to (x_c^-,1^-),
\end{equation}
where $F(s)$ is a {\it scaling function} of combined argument
$s=(x_c-x)(1-q)^{-\phi}$, commonly assumed to be regular at the origin, 
and $\theta$ and $\phi$ are {\it critical exponents}.
The singular behaviour about $q=1$ at the critical point $x_c$ is then given by
$G^{(sing)}(x_c,1)\sim (1-q)^\theta F(0)$.
This scaling assumption implies an asymptotic expansion of the scaling
function of the form
\begin{equation}
F(s) = \sum_{k=0}^\infty \frac{f_k}{s^{(k-\theta)/\phi}}.
\end{equation}
The leading asymptotic behaviour characterises the singularity of the
perimeter generating function via $G(x,1) \sim f_0 (x_c-x)^{-\gamma}$,
where $\theta+\phi\gamma=0$.
The first singularity of $F(s)$ on the negative axis determines the
singularity along the curve $x_c(q)$.
The locus on the axis (say at $s=s_c$) determines the line $x_c(q)\sim
x_c-s_c(1-q)^\phi$ near $q=1$, which meets the line $q=1$ vertically
for $\phi<1$.

The area moments $g_k(x)$ are defined as coefficients of the (formal)
expansion of the generating function about $q=1$
\begin{equation}\label{form:armom}
G(x,q) = \sum_k g_k(x) (1-q)^k, \qquad
g_k(x) = \frac{(-1)^k}{k!} \sum_{m,n} n(n-1)\cdots(n-k+1)p_{m,n} x^m.
\end{equation}
Note that $g_0(x)=G(x,1)$ is the perimeter generating function.
The higher moments $g_k(x)$ can be computed recursively using
$g_l(x)$, where $l<k$ using the functional equation \cite{R02}.
There is an important relation between the scaling function and the
area moments:
The coefficients $f_k$ in the asymptotic expansion of the scaling
function are the leading singular amplitudes of the area moments
$g_k(x)$.
This follows from the scaling assumption which implies that the
leading singular behaviour of the area moments is of the form
\begin{equation}\label{form:gsing}
g_k^{(sing)}(x) \sim \frac{f_k}{(x_c-x)^{\gamma_k}} \qquad (x\to x_c^-),
\end{equation}
where $\gamma_k=(k-\theta)/\phi$.
Thus, information about the scaling function can be obtained from the
leading singular behaviour of the area moments which are, in turn,
obtainable by a numerical analysis, extrapolating finite-size data. 

\section{$q$-Algebraic functional equations}

Due to the difficulty of the SAP problem, simple subclasses of SAPs
have been analysed in the past; for example rectangles, Ferrers
diagrams, stacks, staircase polygons, bar-graph polygons, convex
polygons and column-convex polygons, among others
\cite{Bou96,BG90b,PB95,PO95b}.
All these models share the special property that they are directed,
i.e. they can be thought of as being built up by adding consecutive
layers of squares.
This results in a special recursion for the coefficients $p_{m,n}$ or
a specific functional equation for the generating function $G(x,q)$.
Over the years it has become clear that the underlying structure is a
{\it $q$-algebraic functional equation} of the form
\begin{equation}\label{form:qalg}
P(G(x,q), G(qx,q), \ldots, G(q^Nx,q),x,q)=0,
\end{equation}
where $P(y_0,y_1,\ldots,y_N,x,q)$ is a polynomial in
$y_0,y_1,\ldots,y_N,x$ and $q$.
All exactly solved polygon models satisfy such a functional equation
(with $x$ conjugate to the (horizontal) perimeter and $q$ conjugate to
the area).
For the example of bar-graph polygons, see the next section.
No such recursion is known for SAPs, however.
The limit $q\to 1$ in (\ref{form:qalg}) 
generally leads to an {\it algebraic} differential equation,
i.e. an algebraic equation in $x$, $G(x,1)$ and its derivatives.
The exactly solved polygon models, however, result in a purely
algebraic equation, i.e. in the non-trivial polynomial identity
\begin{equation}\label{form:alg}
P(G(x,1), G(x,1), \ldots, G(x,1),x,1)=0.
\end{equation}

Not much is known about the singular behaviour of general
$q$-algebraic functional equations.
The $q$-linear case, however, has been treated in some detail, see
\cite{RG01} and references therein.
The polygon models of rectangles, Ferrers diagrams, and stacks are
$q$-linear \cite{PO95a}, whereas the other exactly solved models cited
above satisfy $q$-quadratic functional equations.
Using (\ref{form:alg}), we see that the perimeter generating function
$G(x,1)$ displays an algebraic singularity at some point $x_c$. 
The most direct way to extract the scaling function about $x=x_c$ and
$q=1$, whose existence we assume, from the defining $q$-functional
equation, is to insert the Ansatz (\ref{form:scaling}) into the
functional equation, introduce the scaling variable
$s=(x_c-x)(1-q)^{-\phi}$ and expand the functional equation to leading
order in $\epsilon=1-q$.
This will, to lowest order in $\epsilon$, result in a differential
equation for the scaling function $F(s)$.
Above method is also called method of {\it dominant balance} \cite{PB95}.
Note first that the expansion of the generating function with
$q$-shifted argument is given to leading order by
\begin{equation}
G(q^k x,q) = G^{(reg)}(x,q) + \epsilon^{\theta} F(s) 
+ k x_c \epsilon^{\theta+ (1-\phi)} F'( s) + 
{\cal O}( \epsilon^{\theta+ 2(1-\phi)}).
\end{equation}
We will analyse the case of a square-root singularity of the perimeter generating function in detail.
(The case of a simple pole is discussed in \cite{R02}).
It imposes the restrictions
\begin{equation}
P = 0, \qquad 
\left( \sum_k \partial_k P \right) = 0, \qquad 
\left( \sum_{k,l}\partial_{k,l} P \right) \neq 0, \qquad 
\left( \partial_x P \right) \neq 0
\end{equation}
on the functional equation, evaluated at $(x,q)=(x_c,1)$.
We introduced the abbreviations $\partial_k=\partial_{y_k}$ and 
$\partial_{k,l}=\partial_{y_k}\partial_{y_l}$ for $0\le k,l \le N$.
The assumption of a square-root singularity also implies
$\phi=2\theta$ for the critical exponents, as follows from the
relation $\theta+\phi\gamma=0$.
The functional equation then has an expansion of the form
\begin{equation}
0 = \epsilon^{2\theta} \frac{1}{2}  \left( \sum_{k,l} \partial_{k,l} P
\right)  F^2(s)
+ \epsilon^{\theta+(1-\phi)}  x_c \left(\sum_k k \partial_k P \right) F'(s) 
-  \epsilon^{\phi} \left( \partial_x P \right) s
+ \cdots
\end{equation}
to lowest orders in $\epsilon$.
In order to obtain a non-trivial equation for the scaling functions,
we demand equality of the coefficients.
This results in exponents $\theta=1/3$ and $\phi=2/3$, and we get the Riccati
equation
\begin{equation}\label{form:Riccati}
F(s)^2 - 4 f_1 F'(s) - f_0^2 s = 0,
\end{equation}
where the coefficients $f_0$ and $f_1$ are given by
\begin{equation}
f_0^2 = \frac{\left( \partial_x P \right)}{\frac{1}{2}\left( \sum_{k,l}
\partial_{k,l} P \right)},  \qquad
- 4 f_1 = x_c \frac{\left(\sum_k k \partial_k P \right)}
{ \frac{1}{2}\left(\sum_{k,l} \partial_{k,l} P\right)}.  
\end{equation}
The equation, together with the prescribed asymptotic behaviour, has
the unique solution
\begin{equation}\label{form:logairy}
F(s) = -4 f_1 \frac{d}{ds} \ln \mbox{Ai} \left( \left(\frac{f_0}{4
f_1}\right)^{2/3}s\right),
\end{equation}
where $\mbox{Ai}(x)=\frac{1}{\pi}\int_0^\infty\cos(t^3/3+tx) \, dt$ is
the Airy function.
The scaling function is regular at the origin, and its first
singularity on the negative real axis is a simple pole, located at
$\left( \frac{f_0}{4f_1}\right)^{2/3}\!\!s_c=-2.338107...$ .

It is important to notice that the functional form of the result is
{\it independent} of the detailed form of the functional equation, as
long as the coefficients $f_0$ and $f_1$ are nonzero and finite!
Thus, the scaling behaviour is determined by the type of singularity
of the perimeter generating function.
This remarkable kind of universality allows us to conjecture the
functional form of the scaling function for self-avoiding polygons:
Together with the (numerical) observation that the perimeter
generating function of the model of $\it rooted$ SAPs
$G^{(r)}(x,q)=x\frac{d}{dx}G(x,q)$ also has a square-root singularity,
one might ask if it has the scaling function (\ref{form:logairy}).

\section{Bar-graph polygons}

\begin{figure}[htb]
\begin{center}
\begin{minipage}[b]{0.95\textwidth}
\center{\epsfig{file=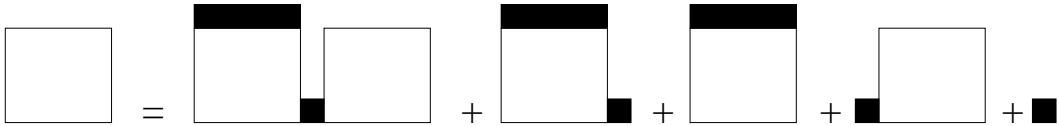,width=14cm}}
\end{minipage}
\end{center}
\caption{\label{fig:barrek}
\small Diagrammatic representation of the bar-graph polygons
functional equation \cite{PB95}.}
\end{figure}

The model of bar-graph polygons has been defined and solved exactly in
\cite{PB95}.
We review the derivation and extract the scaling behaviour as outlined above.
Let $G(x,y,q)$ denote the anisotropic perimeter and area generating
function for bar-graph polygons.
The variable $q$ is conjugate to the area, while the variables $x$ and
$y$ are conjugate to the horizontal and vertical perimeters respectively.
By partitioning the set of bar-graph polygons into subsets of inflated
and concatenated ones, it is straightforward to show that $G(x,y,q)$
satisfies a $q$-quadratic functional equation in the horizontal
perimeter variable $x$ and the area variable $q$
\begin{equation}
G(x,y,q) =  G(qx,y,q) \,qx \, G(x,y,q)+y\, G(qx,y,q)\, qx+y \,
G(qx,y,q)+qx\, G(x,y,q)+y\,qx.
\end{equation}
This is depicted in Figure \ref{fig:barrek}.
The $q$-quadratic functional equation can -- similarly to the
treatment of a Riccati differential equation -- be linearised by means
of the transformation
\begin{equation}
G(x,y,q) = \frac{y}{qx}\left( \frac{H(qx,y,q)}{H(x,y,q)}-(1+qx)\right),
\end{equation}
and the resulting $q$-linear functional equation of order two for
$H(x,y,q)$ can be solved by iteration.
This leads to
\begin{equation}
H(x,y,q) = \sum_{n=0}^\infty \frac{(-qx(1-y))^n
q^{\binom{n}{2}}}{(q;q)_n (y;q)_n},
\end{equation}
where $(t;q)_n=\prod_{k=0}^{n-1}(1-tq^k)$ denotes the $q$-product.
The appearance of a $q$-series is a generic feature of solutions of
$q$-algebraic functional equations, making it difficult to extract the
asymptotic behaviour for $q\to 1$ directly from an exact solution.
For staircase polygons, the first exactly solved model with a
$q$-quadratic functional equation \cite{BG90b,PB95}, this has been
done by use of saddle point techniques \cite{P95}, and this method can
also be applied to the above solution.
A more direct approach consists in deriving the (anisotropic) scaling
function directly from the defining functional equation by use of the
above methods.
To this end, note that the defining polynomial is of the form
$P(z_0,z_1,x,q)=qxz_1z_0+(1+qx)yz_1+(qx-1)z_0+qxy$.
The perimeter generating function may be found by setting $q=1$ and
solving the functional equation for $G(x,1)$. 
It has a square-root singularity at $x_c(y)=\frac{1+y-2\sqrt{y}}{1-y}$.
An analysis along the above lines yields
\begin{equation}
f_{0}^2 = \frac{(1-y)^2(1-x_c(y)^2)}{4\,x_c(y)^3},
\qquad -4 f_1=\frac{1}{2}\left(1+y-(1-y)\,x_c(y)\right),
\end{equation}
and the scaling function is the logarithmic derivative of an Airy
function, given by (\ref{form:logairy}).
The critical exponents are $\theta=1/3$ and $\phi=2/3$.

\section{Self-avoiding polygons}

The model of {\it rooted} SAPs, $G^{(r)}(x,q)=x\frac{d}{dx}G(x,q)$,
has been analysed numerically some time ago, where it was found to
exhibit the same exponents $\theta=1/3$ and $\phi=2/3$ as bar-graph
(staircase) polygons which, in particular, implies a square-root
singularity of the perimeter generating function (see also
\cite{J00}).
With the improved series data now at hand and the refined techniques
of analysis, we were able to confirm our conjecture about the scaling
function to high precision using data obtained from exact enumeration
for small values of perimeter, on the square lattice and on the
triangular lattice \cite{RGJ01}.
In addition, it allowed us to analyse the first two corrections to
scaling \cite{R02}.
Here, we present results obtained from an analysis of extended
enumeration data for square, hexagonal and triangular lattices and in addition
we present data for the square lattice obtained from Monte-Carlo simulations.
The underlying assumption is that the scaling behaviour of rooted SAPs may be 
arbitrarily well approximated by the scaling behaviour of solutions to
a $q$-algebraic functional equation with a square-root singularity.
The scaling function (\ref{form:logairy}) implies certain relations
between the amplitudes $f_k$ of the leading singularity
(\ref{form:gsing}) of the area moment functions $g_k(x)$.
Using the Riccati equation (\ref{form:Riccati}) for the scaling function, 
it can be shown that $f_n = c_n f_1^n f_0^{1-n}$, 
where the numbers $c_n$ are given by the quadratic recursion 
\begin{equation}\label{form:c}
c_n + (3n-4) c_{n-1} + \frac{1}{2}\sum_{r=1}^{n-1} c_{n-r}c_r=0, \qquad c_0=1.
\end{equation}
The first few values are
$c_1=1$, $c_2=-5/2$,
$c_3= 15$, $c_4=-1105/8$, $c_5=1695$, $c_6=-414125/16$, $c_7=472200$, 
$c_8=-1282031525/128$, $c_9=242183775$, $c_{10}=-1683480621875/256$.
The coefficients $f_k$ may be estimated from series enumeration data:
Assuming  an asymptotic growth of coefficients of the area moment
functions $g_k(x)$ of rooted SAPs of the form
\begin{equation}\label{form:fsize}
[x^m] g_k(x) \sim (-1)^k E_k x_c^{-m} m^{\gamma_k-1} \qquad (m\to\infty)
\end{equation}
implies a singular behaviour of $g_k(x)$ of the form (\ref{form:gsing}) with
$f_k=(-1)^k \frac{E_k}{\sigma} x_c^{\gamma_k}\Gamma(\gamma_k)$ and
$\gamma_k=(3k-1)/2$.
The constant $\sigma$ is defined such that $p_{m,n}$ is nonzero only
if $m$ is divisible by $\sigma$.
Thus $\sigma=2$ for the square and hexagonal lattices and $\sigma=1$
for the triangular lattice.
Note that the amplitudes $E_k$, defined by (\ref{form:fsize}), take
the same values for rooted SAPs and for (ordinary) SAPs.
Extrapolation techniques allow us to estimate $x_c$, $\gamma_k$ and $E_k$ to
high precision.
This route was followed in \cite{RGJ01,R02} where we used series data
from exact enumeration up to perimeter 86 on the square lattice and up
to perimeter 29 on the triangular lattice, where the same behaviour
(\ref{form:fsize}) is expected due to universality. 

Since then the enumeration of area-moments for square lattice SAPs has
been extended to perimeter 100 (110) \cite{IJ03a}, and series have
been derived for the hexagonal lattice to perimeter 140 (156) and the
triangular lattice to perimeter 58 (60) \cite{IJ03b}, where the
numbers in parenthesis refer to slightly longer series calculated for
the perimeter generating function. 
Details of analysis can be found in \cite{JG99,J00} and a brief
summary will suffice here.
Estimates for the critical points and exponents are obtained using the
numerical technique of differential approximants \cite{G89}. For the
square and hexagonal lattices, where $\sigma=2$ and only coefficients
with $m$ even are non-zero, we actually analyse the functions
$h_k(y)=\sum_{m\geq 0,n} n^k p_{2m+j,n}y^m$, where $j=4$ and 6 is the
perimeter of the smallest SAP on the square and hexagonal lattices,
respectively.
A singularity at $x=x_c$ in $g_k(x)$ thus becomes a singularity at
$y=x_c^2$ in $h_k(y)$.
The generating functions for the square and triangular lattices show
no convincing signs of singularities other than the one at $x_c$. In
these cases we fit the coefficients to the assumed form
\begin{equation}\label{form:momampl}
\sum_n n^k p_{m,n}
\sim x_c^{-m} m^{\gamma_k-1} k! [E_k+\sum_{i\ge 0}a_i/m^{1+i/2}]
\qquad (m\to\infty).
\end{equation}
In the hexagonal case $h_k(y)$ has an additional singularity on the
negative axis at $y=-x_-^2=-0.412305(5)$ with exponents which appear
to equal the exponents at $y=x_c^2=0.2928932\ldots$, as first noted in
\cite{EG89}.
Since $x_- > x_c$ the singularity at $y=-x_-^2$ is exponentially
suppressed as $m \to \infty$, however for small values of $m$ it still
has a significant influence on the coefficients in the generating
functions and we have to add an extra sequence to the fit similar to
(\ref{form:momampl}):
\begin{displaymath}
 (-1)^{m/2}|x_-|^{-m} m^{\gamma_k-1} k! [E_- + \sum_{i\ge 0}b_i/m^{1+i/2}].
\end{displaymath}
We obtain several data sets by varying the number of terms used in the fit. 
Only the even coefficients are used in these fits for square and
hexagonal SAPs.
To obtain the final estimates we do a simple linear regression on the data
for the amplitudes as a function of $1/m$ extrapolating to $1/m \rightarrow 0$.
We estimate the error in the amplitude from the spread among the
different data sets.
In this way, we obtain the results for the amplitude combinations listed 
in Table~\ref{tab:momampl}, which have been shown to be independent of
the underlying lattice \cite{CG93,RGJ01}.
The scaling function prediction for these numbers is 
\begin{equation}
E_{2k}E_0^{2k-1} = -\frac{c_{2k}}{4\pi^{3k}} \frac{(3k-2)!}{(6k-3)!}, \qquad
E_{2k+1}E_0^{2k} = \frac{c_{2k+1}}{(3k)!\pi^{3k+1}2^{6k+2}}
\end{equation}
for $k\in\mathbb N$,
where we used the known result that $E_1=\frac{1}{4\pi}$ \cite{C94},
and the numbers $c_k$ are defined in (\ref{form:c}).
It is clear that the estimates for the first 10 area 
weighted moments are in perfect agreement with the predicted exact values.

\begin{table}
\begin{center}
\small
\begin{tabular}{lllll}
\hline \hline
Amplitude  & Exact value &  Square & Hexagonal & Triangular \\
\hline
$E_0$  & unknown & 0.56230130(2) & 1.27192995(10) & 0.2639393(2) \\
$E_1$     & $0.7957747\times 10^{-1}$  & $0.795773(2)\times 10^{-1}$ 
              & $0.795779(5)\times 10^{-1}$  & $0.795765(10)\times 10^{-1}$ \\
$E_2E_0$    & $0.3359535\times 10^{-2}$  & $0.335952(2)\times 10^{-2}$ 
              & $0.335957(6)\times 10^{-2}$  & $0.335947(5)\times 10^{-2}$ \\
$E_3E_0^2$  & $0.1002537\times 10^{-3}$  & $0.100253(1)\times 10^{-3}$ 
              & $0.100255(3)\times 10^{-3}$  & $0.100251(4)\times 10^{-3}$ \\
$E_4E_0^3$  & $0.2375534\times 10^{-5}$  & $0.237552(2)\times 10^{-5}$ 
              & $0.237557(7)\times 10^{-5}$  & $0.237547(6)\times 10^{-5}$ \\
$E_5E_0^4$  & $0.4757383\times 10^{-7}$  & $0.475736(3)\times 10^{-7}$ 
              & $0.475749(10)\times 10^{-7}$  & $0.475724(15)\times 10^{-7}$ \\
$E_6E_0^5$  & $0.8366302\times 10^{-9}$   & $0.836624(5)\times 10^{-9}$ 
              & $0.836652(10)\times 10^{-9}$  & $0.83660(2)\times 10^{-9}$ \\
$E_7E_0^6$  & $0.1325148\times 10^{-10}$  & $0.132514(2)\times 10^{-10}$
              & $0.132519(5)\times 10^{-10}$  & $0.132511(5)\times 10^{-10}$ \\
$E_8E_0^7$  & $0.1924196\times 10^{-12}$  & $0.192418(2)\times 10^{-12}$
              & $0.192426(8)\times 10^{-12}$  & $0.192419(8)\times 10^{-12}$ \\
$E_9E_0^8$  & $0.2594656\times 10^{-14}$  & $0.259464(2)\times 10^{-14}$
              & $0.259472(12)\times 10^{-14}$ & $0.25948(4)\times 10^{-14}$ \\
$E_{10}E_0^9$ & $0.3280633\times 10^{-16}$  & $0.328062(4)\times 10^{-16}$
              & $0.328051(15)\times 10^{-16}$ & $0.32812(5)\times 10^{-16}$ \\
\hline \hline
\end{tabular}
\end{center}
\caption{\label{tab:momampl}
Predicted exact values  for universal amplitude combinations and estimates
from enumeration data for square, hexagonal and triangular lattice polygons.}
\end{table}

\begin{figure}[htb]
\begin{center}
\begin{minipage}[b]{0.95\textwidth}
\center{\epsfig{file=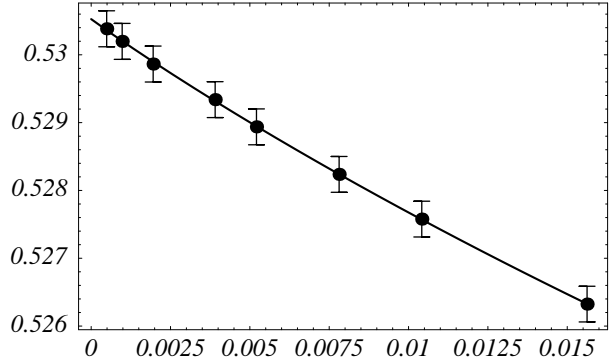,width=8cm}}
\end{minipage}
\end{center}
\caption{\label{fig:D2data}
\small MC estimates of $\langle a^2 \rangle_m/ (2 \langle a
\rangle_m^2)$ plotted against $1/m$.
The estimated statistical relative error is one part in 1000.}
\end{figure}

We also extended our previous approach by analysing SAPs up to perimeter 2048 
by a Monte-Carlo simulation.
An effective method of Monte-Carlo SAP generation of a given perimeter
$m$ has been described in \cite{MOS90}.
Each MC step consists of an inversion or a certain reflection of a
randomly chosen part of the SAP.
Polygons which are no longer self-avoiding are rejected.
The check for self-avoidance (by marking the corresponding lattice
points) as well as the area determination (``on the fly'') can be done
in time proportional to the polygon length.
For a given perimeter $m$, we took a sample of $10^6$ polygons, with
at least $10\times m$ MC update moves between consecutive
measurements.
This leads to an estimated statistical error of 1 part in 1000.
For perimeter 2048, this took two days of CPU time on a 1.5 GHz PC.
We used the random number generator {\tt ran2} described in \cite{PTVF92}.
From the Monte-Carlo measurements of polygons of a given perimeter
$m$, we extracted ratios $D_k/D_1^k$ of the mean area moment
amplitudes $D_k$ defined by
\begin{equation}
\langle a^k \rangle_m = \frac{\sum_n n^k {\tilde p}_{m,n}}{\sum_n
{\tilde p}_{m,n}} 
\sim D_k k! m^{2k\nu} \qquad (m \to \infty),
\end{equation}
where  ${\tilde p}_{m,n}$ is the number of sampled polygons of area $n$,
and $\nu=3/4$ follows from (\ref{form:fsize}).
Note that by using MC sampling we cannot measure the coefficients $E_k$ of 
(\ref{form:fsize}) directly but only the ratios $D_k=E_k/E_0$.
We tested the accuracy of the random number generator by comparing values for 
the area moments obtained by simulation with their exact values, which are
available up to perimeter 100 from exact enumeration.
Assuming the above scaling function results, for $k\in\mathbb N$, in
the values for the amplitude ratios
\begin{equation}
D_{2k}/D_1^{2k} = -c_{2k} \frac{(3k-2)!2^{4k-2}}{(6k-3)!\pi^k}, \qquad
D_{2k+1}/D_1^{2k+1} = c_{2k+1} \frac{1}{(3k)!2^{2k}\pi^k},
\end{equation}
where the numbers $c_k$ are defined in (\ref{form:c}).
We computed the values $\langle a^k \rangle_m/(k!\langle a \rangle_m^k)$ for 
$m=64,96,128,192,256,512,1024,2048$, and a typical data plot is shown
in Figure \ref{fig:D2data}.

We then extrapolated the numbers $D_k/D_1^k$ by a least squares fit to 
$\langle a^k \rangle_m/(k!\langle a \rangle_m^k) \sim
D_k/D_1^k+c_1m^{-1}+c_2m^{-3/2}$ as $m\to\infty$, in analogy with the
series extrapolation for exact enumeration data.
The corresponding results for the first ten amplitude combinations are
given in Table \ref{tab:data}, together with their theoretical values.
The relative error is less than  1 part in 1000.

\begin{table}
\begin{center}
\begin{tabular}{ccc}
\hline \hline
Amplitude & Exact value & MC data\\
\hline
$D_2/D_1^2$ &
0.530516 $\times 10^{-0}$  & 
0.530526 $\times 10^{-0}$ \\
$D_3/D_1^3$ & 
0.198944$\times 10^{-0}$ &
0.198954$\times 10^{-0}$\\
$D_4/D_1^4$ & 
0.592380$\times 10^{-1}$& 
0.592445$\times 10^{-1}$\\
$D_5/D_1^5$ & 
0.149079$\times 10^{-1}$ &  
0.149105$\times 10^{-1}$\\
$D_6/D_1^6$ &
0.329452$\times 10^{-2}$&
0.329534$\times 10^{-2}$ \\
$D_7/D_1^7$ &
0.655743$\times 10^{-3}$ & 
0.655959$\times 10^{-3}$\\
$D_8/D_1^8$ &
0.119654$\times 10^{-3}$& 
0.119705$\times 10^{-3}$\\
$D_9/D_1^9$ & 
0.202754$\times 10^{-4}$ & 
0.202863$\times 10^{-4}$ \\
$D_{10}/D_1^{10}$ &
0.322149$\times 10^{-5}$ & 
0.322376$\times 10^{-5}$ \\
\hline \hline
\end{tabular}
\end{center}
\caption{\label{tab:data} Comparison of prediction for amplitude
ratios against MC data.
The estimated statistical relative error is one part in 1000.}
\end{table} 
The value of $D_1$ is estimated to be $D_1\approx 0.141776$.
We can compare this to our previous analysis by noting that 
$D_1=E_1/E_0\approx 0.14152105(1)$, where $E_1=\frac{1}{4\pi}$ has
been derived by conformal field theory arguments \cite{CG93}, and
$E_0=0.56230130(2)$ together with $x_c=0.379052277757(5)$ has been
determined in \cite{JG99,J00,IJ03a}.
Noting that $f_0=-2\frac{E_0}{\sigma}\sqrt{\frac{\pi}{x_c}}$ and 
$f_1=-x_c \frac{E_1}{\sigma}=-\frac{x_c}{4\pi\sigma}$ we arrive at the
conjectured form of the scaling function for rooted SAPs
\begin{equation}
F^{(r)}(s) = \frac{x_c}{\pi\sigma}\frac{d}{ds}\log\mbox{Ai} \left(
\frac{\pi}{x_c} \left( 2E_0\right)^\frac{2}{3} s \right)
\end{equation}
with exponents $\theta=1/3$ and $\phi=2/3$. 
(Formula (22) in \cite{RGJ01} is correct up to a minus sign.)
The conjectured form of the scaling function is then obtained by
integration and is
\begin{equation}
F(s) = -\frac{1}{\pi\sigma} \log\mbox{Ai} \left( \frac{\pi}{x_c}
\left( 2E_0\right)^\frac{2}{3} s \right)
\end{equation}
with exponents $\theta=1$ and $\phi=2/3$. 
The parameters for the hexagonal lattice are $\sigma = 2$ and
$x_c=1/\sqrt{(2+\sqrt{2})}$ (known exactly from the work of Nienhuis
\cite{N82}) and for the triangular lattice $\sigma=1$ and
$x_c=0.2409175745(3)$.

\section{Conclusion}

We discussed the scaling behaviour of $q$-algebraic functional equations
which underlie the critical behaviour of exactly solvable planar
polygon models, counted by perimeter and area.
This led to a prediction of the scaling function of the unsolved model
of self-avoiding polygons on the square lattice, which we verified to
numerical precision.
The prediction also implies certain amplitude combinations for all
area moments.
It was previously argued \cite{CG93} that these combinations should be
universal, i.e., independent of the underlying lattice. 
Our investigations on the triangular and on the hexagonal lattices
support the universality hypothesis, see also \cite{RGJ01}.
It is possible to extend the above analysis to include the first few
corrections to scaling \cite{R02}.

The question arises whether the SAP area statistics can be found in
other models as well.
One candidate is the hull of planar Brownian motion, which was
observed in 1984 to have the same fractal dimension as SAPs \cite{M83}.
Interestingly, this observation has been proved recently using methods of
stochastic processes \cite{LSW01}.
It might thus be possible to derive the area distribution rigorously for this 
model, as well as a corresponding prediction for the SAP area distribution.

Since $q$-algebraic functional equations are the underlying
mathematical structure of (exactly solvable) planar polygon models
counted by perimeter and area, they may serve to classify possible
scaling behaviour of these polygon models.
$q$-algebraic functional equations exhibit various types of scaling
behaviour, depending on the specific type of singularity of the perimeter
and area generating function \cite{R02}.
Since the techniques applied so far only make use of formal power
series expansions, one might ask for a proof of existence of a uniform
asymptotic expansion.
Details of a proof may rely on techniques used in the related problem of the
behaviour of differential equations containing a parameter \cite{Ol58}
about singular points.
Recently, the scaling function prediction has been rederived using
field-theoretic methods and generalised to higher order critical
points \cite{C01}.
It would be interesting to see whether the proposed scaling functions
of \cite{C01} arise in this framework as well.
It may be possible that $q$-algebraic functional equations also
provide insight into models of polygons with interaction, $q$ then
being the activity of interaction.
There is at least one such exactly solvable model: self-interacting
partially directed walks \cite{OPB93}.

Another possible application might be to models of cluster hulls, which appear
in percolation or in spin models as boundaries of spin clusters in the
Ising model or in the Potts model, as described in \cite{CZ02}.

\section*{Acknowledgements}

CR would like to acknowledge financial support by the German Research
Council (DFG).
IJ and AJG would like to acknowledge the Australian Research Council (ARC) and
IJ would also like to thank the Australian Partnership for Advanced
Computing (APAC) and the Victorian Partnership for Advanced Computing
(VPAC) for generous allocations of computational resources.

\end{document}